\documentclass[10pt,conference]{IEEEtran}
\IEEEoverridecommandlockouts

\usepackage{adjustbox}
\usepackage{xspace}
\usepackage{tabularx}
\usepackage{colortbl}
\usepackage{subfig}
\usepackage{multirow}
\usepackage{booktabs}
\usepackage{listings}
\usepackage{scratch3}
\usepackage[capitalise]{cleveref}
\usepackage{arydshln}
\usepackage{cite}
\usepackage{amssymb}
\usepackage{flushend}
\usepackage{balance}

\setscratch{scale=.60}

\newcommand\definetool[2]{\newcommand{#1}{{\textsc{#2}}\xspace}}
\definetool{\Scratch}{Scratch}
\definetool{\leila}{LeILa}
\definetool{\whisker}{Whisker}
\definetool{\litterbox}{LitterBox}
\definetool{\bastet}{Bastet}
\definetool{\scratchblocks}{scratchblocks}

\colorlet{punct}{red!60!black}
\definecolor{background}{HTML}{EEEEEE}
\definecolor{delim}{RGB}{20,105,176}
\colorlet{numb}{magenta!60!black}

\lstdefinelanguage{json}{
    basicstyle=\normalfont\ttfamily,
    numbers=left,
    numberstyle=\scriptsize,
    stepnumber=1,
    numbersep=8pt,
    showstringspaces=false,
    breaklines=true,
    frame=lines,
    backgroundcolor=\color{background},
    literate=
     *{0}{{{\color{numb}0}}}{1}
      {1}{{{\color{numb}1}}}{1}
      {2}{{{\color{numb}2}}}{1}
      {3}{{{\color{numb}3}}}{1}
      {4}{{{\color{numb}4}}}{1}
      {5}{{{\color{numb}5}}}{1}
      {6}{{{\color{numb}6}}}{1}
      {7}{{{\color{numb}7}}}{1}
      {8}{{{\color{numb}8}}}{1}
      {9}{{{\color{numb}9}}}{1}
      {:}{{{\color{punct}{:}}}}{1}
      {,}{{{\color{punct}{,}}}}{1}
      {\{}{{{\color{delim}{\{}}}}{1}
      {\}}{{{\color{delim}{\}}}}}{1}
      {[}{{{\color{delim}{[}}}}{1}
      {]}{{{\color{delim}{]}}}}{1},
}

\usepackage{fbox}
\newcommand{\rqsummary}[2]{
        \vspace{2mm}
        \noindent
        \fbox{%
            \parbox{.97\linewidth}{%
                    \textbf{#1 Summary.}
                #2
            }%
        }%
        \vspace{2mm}
}%

\usepackage[colorinlistoftodos,prependcaption,textsize=tiny]{todonotes}

\title{The ABC of Pair Programming:
 Gender-dependent Attitude, Behavior and Code of Young Learners}

\begin{document}

\author{%
	\IEEEauthorblockN{Isabella Gra{\ss}l}%
	\IEEEauthorblockA{University of Passau\\
	Passau, Germany\\
	isabella.grassl@uni-passau.de}
\and
	\IEEEauthorblockN{Gordon Fraser}%
	\IEEEauthorblockA{University of Passau\\
	Passau, Germany\\
	gordon.fraser@uni-passau.de}
}

\maketitle

\begin{abstract}
  Young learners are increasingly introduced to programming, and one
  of the main challenges for educators is to achieve learning success
  while also creating enthusiasm. As it is particularly difficult to
  achieve this enthusiasm initially in young females, prior work has identified
  gender-specific differences in the programming behavior of young
  learners.
  Since pair programming, which turns programming into a more sociable
  activity, has been proposed as an approach to support programming
  education, in this paper we aim to investigate whether similar
  gender-specific characteristics can also be observed during pair
  programming.
  Therefore, we designed a gender-neutral introductory \Scratch
  programming course tailored for integrating pair programming
  principles, and conducted it with a total of 139 students aged between
  8 and 14 years. To identify gender-dependent differences and
  similarities, we measure the attitude towards programming and the
  course setting, observe the behavior of the students while
  programming, and analyze the code of the programs for different
  gender-combinations.
  Overall, our study demonstrates that pair programming is well suited
  for young learners and results in a positive attitude. While the
  resulting programs are similar in quality and complexity independent
  of gender, differences are evident when it comes to the compliance
  to pair programming roles, the exploration of code, and the creative
  customization of programs.
  These findings contribute to an in-depth understanding of social and
  technical gender specifics of pair programming, and provide
  educators with resources and guidance for implementing
  gender-sensitive pair programming in the classroom.
\end{abstract}


\section{Introduction}
\label{sec:intro}

Computer science (CS), and programming as one core concept of software
engineering (SE), are skills that are in high demand, hence countries
around the world are increasingly introducing CS in
schools~\cite{denner2021, ma2022}. It is important to introduce
programming early~\cite{chaplin2013,iskrenovic-momcilovic2019b,greifenstein2021} to
provide students with their own experience of programming and to
reduce stereotypical beliefs~\cite{jenkins2002,beyer2014, cheryan2015, ying2019b}. This is particularly relevant since there are still 
fewer females in CS, especially in SE~\cite{albusays2021}.

%

%

To address this challenge and to demonstrate to young females that
programming is indeed a social activity which requires communication
and collaboration~\cite{choi2015a, demir2021}, the team-based method
of \emph{pair programming} (PP) has the potential to support females
in learning to program. 
%
In general, PP is a well established agile technique in professional software
development with its distinctive feature of distributing the
programming tasks within two pre-defined roles~\cite{beck1999} of a
\textit{driver}, who does the coding, and a \textit{navigator} who
advises and gives feedback to the driver. These role definitions make
PP well suited for education, since it can be better ensured that both
students in the PP pair participate in programming on an equal
level~\cite{williams2003}.
However, there has been little research on whether and how gender differences occur in PP when applying it to younger programming learners~\cite{iskrenovic-momcilovic2019b}. 

In this paper, we aim to bridge the gap in knowledge about the role of different pair combinations among young novices by identifying gender-dependent characteristics of PP in introductory programming courses.
%
%
We designed an introductory programming in-class course targeting the
integration of basic PP principles into \Scratch, and conducted it
nine times at several schools with 139 students between the ages of 8 and 14 with no
prior programming experience. We emphasized a
gender-neutral course design to reduce potential gender stereotypes~\cite{sullivan2016,grassl2022}. 
%
%
We measure gender-dependent effects on three key factors of relevance
as demonstrated by previous studies: (1) the students' attitude
towards programming and the course design, (2) the behavior within the
team, and (3) the process of programming as well as the resulting code.

Our study demonstrates that our course design successfully introduces
PP to young novice programmers, resulting in a positive attitude
towards programming and the course design. We observe a preference for
the role of driver independently of pair constellation, but all-male
pairs adhere less to the role allocation than all-female pairs, and
are more adventurous in exploring the programming environment. We also
observe gender stereotypical preferences when giving pairs creative
freedom, just like during individual programming, with all-female
pairs focusing more on costumes and backgrounds. Overall, PP can be implemented successfully in lower grades and provides
young students, especially females, a positive initial CS experience,
although teachers have to pay attention to certain gender-dependent
behaviors.
Our findings support educators to design courses that encourage young females to get involved in programming at an earlier stage.

To support replications and future research we provide all course
materials and evaluations for replication
online.\footnote{https://doi.org/10.6084/m9.figshare.21878859.v1 All sensitive data are available upon request.}

\section{Background and Related Work}
\label{sec:background}

\subsection{Learning to Program: \Scratch and Pair Programming}

CS and especially its subfield of SE are often perceived as abstract and male-dominated by society~\cite{beyer2014, jenkins2002, cheryan2015, liebenberg2010, ying2019b}. Although learning to program can be difficult and frustrating~\cite{jenkins2002, peterson1993}, it is important to introduce it already at a young age~\cite{chaplin2013, iskrenovic-momcilovic2019b}. 
The block-based programming environment \Scratch offers an easy and playful introduction to programming for young learners: The use of blocks avoids the necessity to learn syntax, and the many figures (\textit{sprites}) and backgrounds (\textit{stages})~\cite{resnick2009a} offer ample creative potential. 

A common approach in general to improve the programming process is to work in teams. For example, in professional software development PP is often used~\cite{beck1999}. In PP, two developers work on a problem in a pair, using the roles of driver and navigator: While the driver is the person who types code, the navigator is responsible for strategic guidance and providing feedback.
Research has shown that young learners are more successful when programming in pairs rather than alone. In particular, performance and programming knowledge has been found to be higher~\cite{iskrenovic-momcilovic2019b, cal2020, denner2014, papadakis2018b, liebenberg2010, issaee2021}, fun and positive attitude increase~\cite{cal2020, papadakis2018b}, and PP is beneficial for self-confidence and interest~\cite{berenson2004}. 

Besides the immediate advantages for learners, teachers also benefit from students helping each other, thus reducing the supervision effort~\cite{ying2019b, hanks2008}. However, at the same time teachers have to ensure that the rules of PP are clearly communicated to avoid confusion~\cite{williams2008, cal2020} and inhibiting learning success. A further question to consider is how to build pairs, since grouping is often key to success, and students should feel comfortable and have a good learning atmosphere that strengthens social aspects like communication and collaboration~\cite{denner2014, demir2021}.

\subsection{Gender in Programming Education and Pair Programming}

Girls and boys often behave differently in CS education and programming classes in particular: while girls follow teachers' instructions better and are more quiet, boys are often more intrusive and active~\cite{harskamp2008, pomerantz2001, chaplin2013, ma2022, wang2020e}. This is socioculturally inherently learned, with the result that girls are more inclined to please teachers~\cite{liebenberg2010, pomerantz2001}.

Although females in particular benefit from collaboration~\cite{kong2018, williams2003, zhong2016a}, research has so far neglected the question of gender-related characteristics in different pair constellations---all-female, all-male, and mixed pairs---with young students. 
At team level, some studies with teenagers and bachelor students indicate no significant gender-dependent differences between pairs in code production~\cite{demir2021, zhong2016a, gomez2017a, kung2022}, while others highlight that all-male groups have higher quality output~\cite{tsan2016, jiang2017, jarratt2019}. 
The social aspect of PP has been studied less in detail, especially with young learners---the results so far indicate that the group constellation may well have an influence on attitude~\cite{zhan2015b}, communication and behavior~\cite{choi2015a, jiang2017, tsan2021, kung2022}. Since girls often have a negative attitude due to the social image of programming~\cite{sun2022, taylor2019}, the effects of team constellation merit further investigation.

The children themselves also prefer different constellations: While boys tend to like mixed groups, girls prefer group work with other girls~\cite{harskamp2008}. Although there are few studies on this, one assumption behind this observation is that boys can assert leadership and girls simply follow~\cite{harskamp2008, zhan2015b, wieselmann2020a}. 

There are three limitations of prior work which we aim to address with this paper: (1) there have been few studies with children at elementary school and early middle school~\cite{iskrenovic-momcilovic2019b, kung2022}, (2) most prior studies were conducted outside a school context, and thus often few female pairs were studied since mostly boys are interested in extracurricular programming projects~\cite{kung2022, jarratt2019}, and (3) little attention has been paid so far to affective effects such as behavior or creativity of PP roles~\cite{iskrenovic-momcilovic2019b, kung2022, roman-gonzalez2018}.
This gap is problematic since we require a better understanding in how girls in particular benefit from different PP pair constellations, such that they can experience as early as possible that they, too, can program~\cite{iskrenovic-momcilovic2019b}, which in turn might get them interested in CS before socially learned gender stereotypes become more prevalent~\cite{cheryan2015}. 
Therefore, we aim to address these challenges by determining gender-dependent attitudes, behavior, and code of students aged 8 to 14 years old in a regular school environment.

\section{Course Design}
\label{sec:course}

The aim of this paper is to identify gender-dependent differences and similarities among different pair constellations when using PP in elementary and early middle school.
Therefore, we designed and implemented an introductory programming in-class course for \Scratch which is specifically tailored for integrating the basic principles of PP, targeting students without prior CS experience from 8 to 14 years old.

\begin{figure}[tb]
\centering
{\includegraphics[width=\columnwidth]{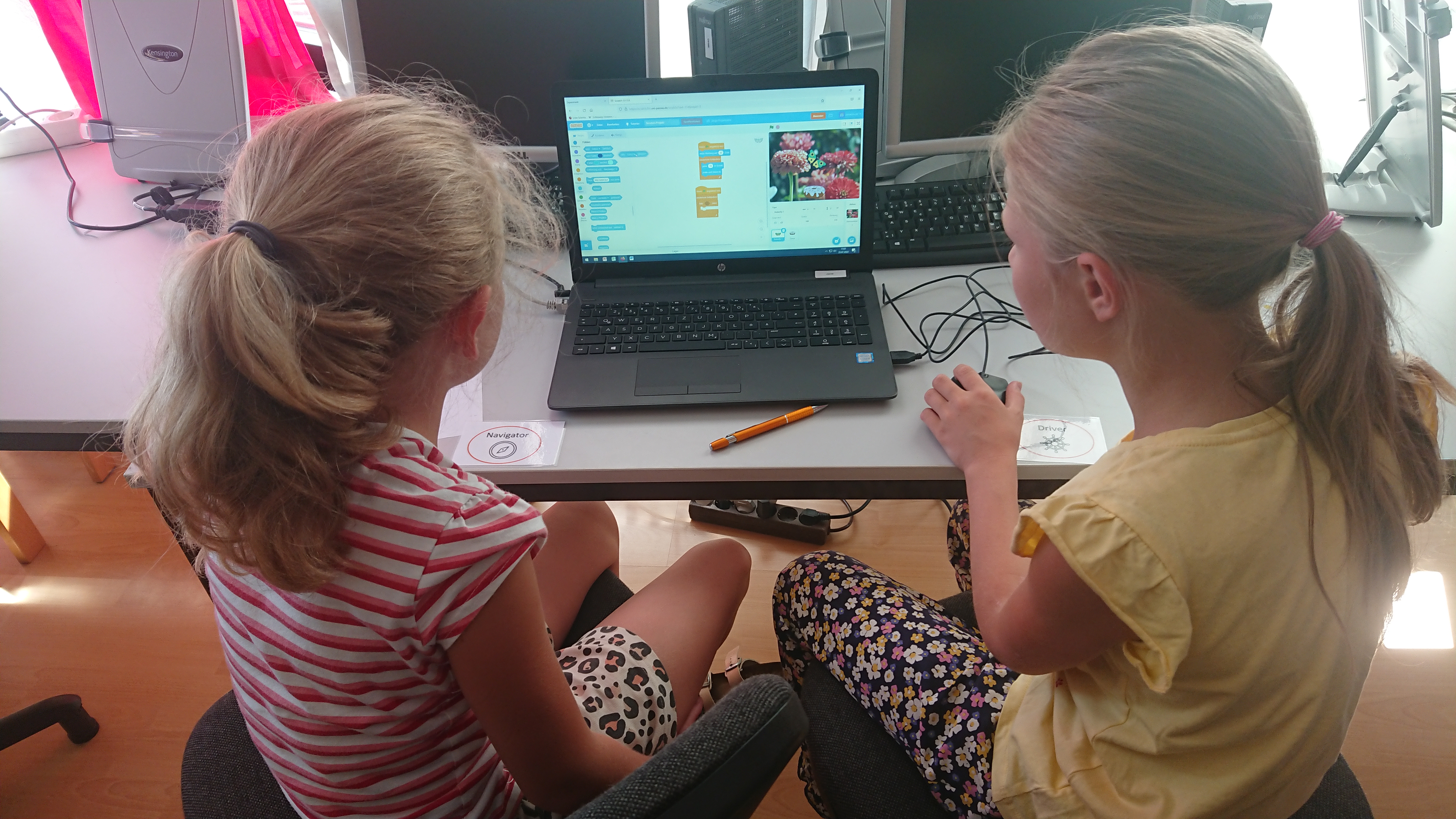}}
\caption{Female students with their physical PP role cards.}
\label{fig:roles}
\end{figure}

\subsection{Introducing Young Learners to Pair Programming}
Since the targeted young learners are not familiar with the concept of PP, the key challenge is to integrate PP into an in-class course to provide them a coherent introduction to it according to the guidelines established by Williams et al.~\cite{williams2008}.

Before the class begins, the set up needs to be prepared by arranging two chairs and a desk together for each pair, with sufficient distance from the next pair, and moving the keyboard, mouse and monitor to the center so that both students in a pair have access~\cite{williams2008}.

To start the introduction, we address the students' lifeworld by asking them where they already experience teamwork in their (school) everyday life as well as which benefits and challenges they experienced.
After a brief discussion, we point out that the distribution of tasks and contributions in teams often remains unclear, which can be frustrating~\cite{wieselmann2020a}. 
In this way, we explain why there exist two different roles in PP, which have different responsibilities, but a common goal~\cite{williams2008}, and that today the students slip into these roles like little actors. 

To keep the course interactive, we ask the students about their ideas of what the roles represent and what the roles are supposed to do. After collecting the answers in a plenary session, we explain the roles using the analogy of sea and car travel, hence, establishing an association to the students' lifeworld while ensuring an understanding of the roles.
To further encourage the students, we assure them that adult computer scientists also use this method.

\subsection{Implementation of Pair Programming}

Since the following of the PP protocol, i.e., adhering to assigned roles, is one of the main challenges of introducing PP to young students~\cite{williams2002a}, we illustrate the roles in an accessible, playful, and age-appropriate way using small cards. Each pair receives two physical cards with the names of the roles and a matching icon for each role~(\Cref{fig:roles}). 
The card of the driver shows an icon of a control board, representing the individual in charge of writing the actual code. The card of the navigator displays an icon of a compass to illustrate the importance of navigating, thinking, observing and giving feedback to the driver~\cite{williams2002a}. 
These cards are distributed so that each student has a card with their role in front of them, and we tell them that they must swap with their partner after each task.

The course is organized in terms of an even number of tasks, with roles changing between each successive task.
The students are reminded that only the driver may access the keyboard and the mouse---just like a steering wheel in transportation, the navigator must not grab the mouse---, and that the roles are switched after each task. In order to enforce the PP protocol, the solution of a task is discussed in plenum after each task, and the students are reminded to swap their cards. Supervisors monitor and ensure that cards are swapped.

Throughout the course, the supervisors encourage the students to find their own solutions for problems rather than telling them the correct solution, since working in pairs supports problem solving without frustration~\cite{williams2008}.


\begin{figure}[tb]
\centering
{\includegraphics[width=\columnwidth]{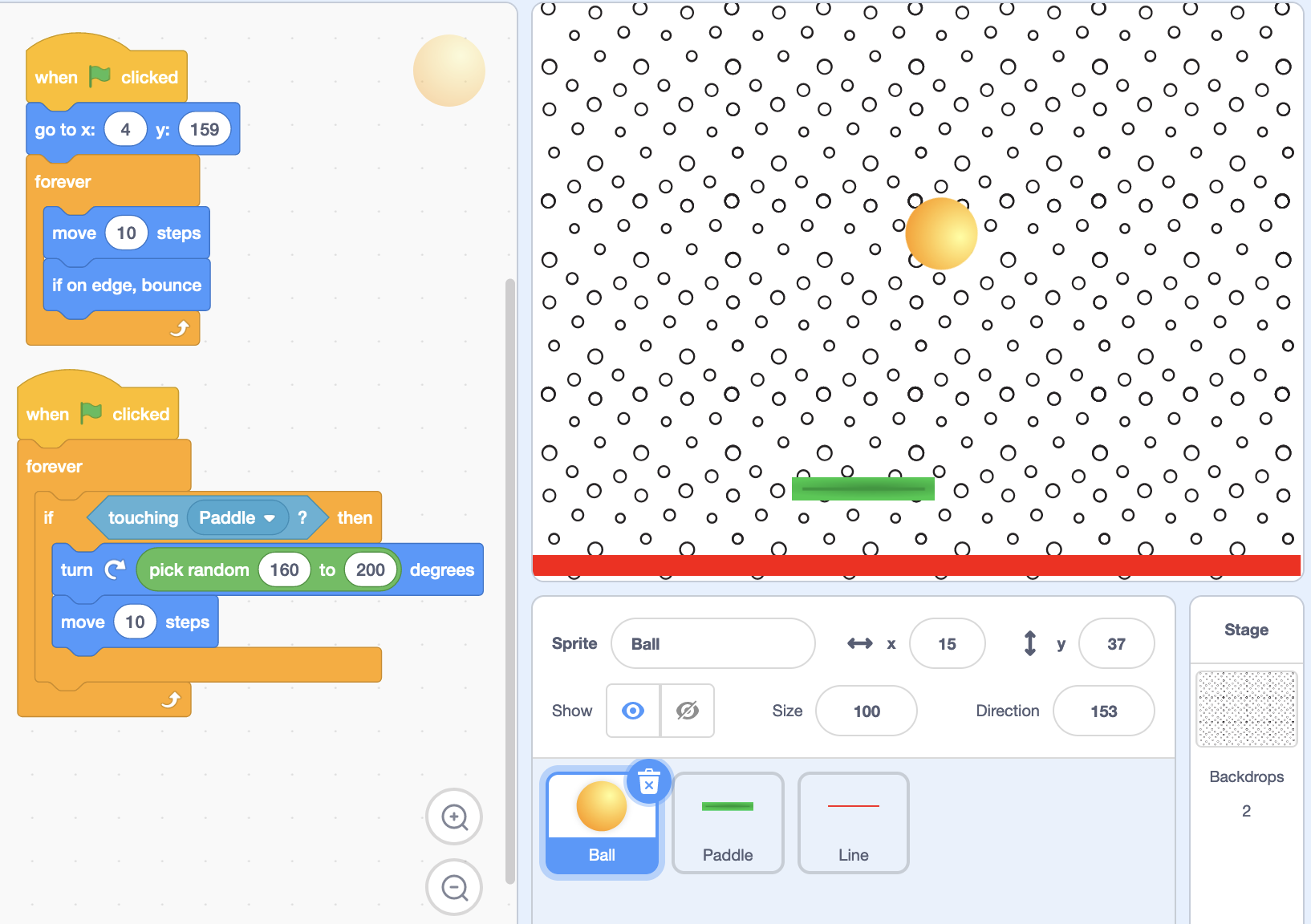}}
\caption{Sample of the game \textit{Pong} used in the course design.}
\label{fig:pong}
\end{figure}

\subsection{Course Schedule}
The aim of the course is the implementation of the well-known game \textit{Pong}~(\cref{fig:pong}), which is also introduced as one of the official tutorials on the website of~\Scratch\footnote{https://scratch.mit.edu/projects/editor/?tutorial=all}. We selected a game as project type to ensure that all students apply the same basic programming concepts, since girls tend to prefer projects types with sequential flow and thus, applying less complex structures~\cite{grassl2021}. 
The course is explicitly designed as an in-class activity, hence, the basic programming concepts shall be introduced and recalled within a manageable time in the curriculum~\cite{buckley2003}. Thus, the course lasts about 1.5 hours and consists of six tasks in order to have an even number of tasks for each pair. We also ensure that there are two creative tasks at the end of the course so that both students in a pair get a chance to realize their ideas, thus balancing individual preferences and collaborative work~\cite{williams2008}.
Furthermore, the difficulty of the tasks increases gradually to ensure that each individual task has a similar scope and that the complexity of the tasks could be reduced in time if necessary. 
The order and selection of programming concepts is based on the principle that sequences represent the simplest structures, followed by loops and conditional statements~\cite{grover2015}.

\subsubsection{Task 1: Basic Instructions} \label{sec:task1}
The first task starts with deleting the default sprite, and encouraging the students to select an individual stage and sprite. 
To maximize creative freedom and to reduce gender stereotypes, we select a neutral stage (\textit{dots}) and a neutral shooting object (\textit{ball}) in our sample~\cite{rubegni2020}.
After each pair has chosen a stage and sprite, the interface of \Scratch and the function of the \textit{green flag} is explained by using an analogy from the student's lifeworld: code in \Scratch needs a trigger, just like videos on \textit{Youtube} need a play button~\cite{adams2012}. Then, we ask the students which of the blocks of the \textit{motion}-category might be useful for moving our sprite. As a result, sequences are created by adding \texttt{move}-blocks to the green flag event handler, and students try out how the ball moves some steps to the right.

\subsubsection{Task 2: Loops} \label{sec:task2}
The programming concept \textit{loops} (\texttt{forever, repeat-times}) is introduced by motivating that we do not want to repeatedly press the green flag, nor do we want to concatenate many motion blocks. The students should discover the \texttt{forever}-block by browsing through the categories. 
By trying it out on their own, they should notice that the ball moves to the right edge and stops. By discussing how to change this, ideally students will come up with the idea that the ball should bounce off the wall, for which they can find a pre-defined block in the \textit{motion}-category in \Scratch. This also familiarizes them further with the interface.

\subsubsection{Task 3: Conditional Statements} \label{sec:task3}
In order to balance programming and being creative~\cite{grover2015, williams2008}, we first have the students pick a new sprite, which they should control with the mouse. Once again, we select a gender-neutral sprite in our sample (\textit{paddle})~\cite{rubegni2020}. 
The player moves the newly selected sprite so that the ball does not touch the bottom. Since the sprite is supposed to follow the mouse only on the abscissa, we briefly explain the coordinate system. 
We introduce the new block category \textit{sensing}, where the mouse is set to the x-position. To make sure that the sprite follows the mouse continuously, we need a \texttt{forever}-block: the students should first find a solution within their pairs and then discuss this in plenary, thus supporting recall of the concept of loops~\cite{buckley2003}.

At this stage of the game, the ball passes through the paddle instead of interacting with it, and thus, the usefulness of the next major programming concept, conditions~(\texttt{if, if-else}), is illustrated. To reduce the complexity of the principle, we use analogies from the students' lifeworld by brainstorming with them, e.g., if it is raining we need an umbrella. 
We give the hint that conditional blocks are elements of the control of \Scratch and let the pairs find the correct block on their own. As soon as every pair has found the correct block, we develop the pseudo code of our program in the plenary, i.e., which condition we need and what should happen then. Afterwards the students try it out themselves in pairs.

\subsubsection{Task 4: Practicing Loops \& Conditionals} \label{sec:task4}
 In order to make the game more exciting and complex, we change the angle of incidence and reflection of the ball when it bounces off the paddle. We introduce operators for this purpose, and the students are asked to experiment with different angles. 
To ensure that the ball always starts at the same point, a \texttt{go to}-block should be added before the loop in the ball. This concludes the basic functionality of the game.
Before continuing with recalling the learned programming concepts, we encourage the students to pick a third sprite, place it on the bottom, and on contact with the ball the game is over. We demonstrate this using the gender-neutral option of the \textit{line}. 
Similar to the code of the ball interacting with the paddle, this sprite requires code including loops and conditions. To enable more in-depth recall~\cite{buckley2003}, we encourage the students to write the code for the line themselves, and discuss the solution together afterwards.

\subsubsection{Task 5 \& 6: Free Task 1 \& 2} \label{sec:task5}
From this point on, we let the students work on their own, and mainly support them in implementing their own ideas. The driver of each task gets to choose what they implement, and the next task is switched, so that both students are able to implement their own ideas in the creative task~\cite{williams2008}.
Before they start, a few ideas for extensions are briefly collected in the plenary such as experimenting with costumes, stages, colors, speed of  objects, levels, etc. We show the students the remaining block categories \textit{looks}, \textit{sound} and \textit{variables} and try to inspire them by stating different extensions such as creating multiple sprites, having a multiplayer mode, introducing levels and scores as well as winning or losing screens. We again encourage the students to be creative~\cite{williams2008, rubegni2020}. The course session closes with the students showing their created projects to the other students.

\section{Method}
\label{sec:method}

Using the PP-based \Scratch course (\cref{sec:course}), we aim to
empirically answer the following research questions:

\begin{itemize}
\item \textbf{RQ1:} \textit{Does the pair constellation in PP relate to the students' attitude towards programming?}
\item \textbf{RQ2:} \textit{Does the pair constellation in PP relate to the students' behavior while programming?}
\item \textbf{RQ3:} \textit{Does the pair constellation in PP relate to the students' programs?}
\end{itemize}

\subsection{Pre-Study}

To evaluate the course design, we conducted a pre-study with 15 students distributed in six pairs and one group of three. The main objective of the pre-study was to evaluate the difficulty of the tasks and the time allocation, so that the final course could be easily completed in a double period class. The responses and projects of the pre-study are not included in the analysis of RQ1--RQ3.
We included eight tasks in the pre-study, which basically had the same content as in our main study. The first two tasks (basics and loops) and the last two tasks (free tasks) are the same in both. 
However, the other tasks were reduced in complexity and combined because of time restrictions: bouncing the ball off the wall was an extra task with conditions, which was then solved with the predefined block of \Scratch and integrated into the second task (\cref{sec:task2}). The task to make the angle of incidence more exciting was an extra task before the free tasks, which also included a more complex formula of operators. The formula was reduced to a minimum of complexity and integrated into the fourth task (\cref{sec:task4}).


\subsection{Data Collection}
\begin{table}[t]
\centering
\caption{Questionnaire for students (\textit{AG-LR}) and supervisors (\textit{CP-HE}) after each task.}
\label{tab:survey}
\begin{tabular}{lll}
\toprule
 Category & Var. & Question  \\ 
 \midrule
 Again & AG&  Would you like to do the task again?  \\
 Fun & FU & How did you like the task?  \\
Like-Role & LR& How did you like your role in the task?  \\
\midrule
 Compliance & CP & The team followed the alloc. of roles in this task. \\
 Collab. & CL& The team collaborated during this task.  \\
 Comm. & CM& The team communicated a lot during this task.  \\
Harmony & HA& The team was harmonious in this task.  \\
Dispute & DI & The team argued during this task.  \\
Help & HE & The team needed help during this task.  \\ 
\bottomrule
\end{tabular}

\end{table}

%

We conducted the final course nine times between April and July 2022, targeting children aged 8 to 14 without prior programming experience. 
An invitation for the course was sent to the schools in the local neighborhood of the University of Passau (Germany) by email. 
The course lasts about 1.5 hours, thus, it is feasible to conduct it within a typical double period class. The pairs grouped themselves either based on the seating in the class or their own preference~\cite{williams2008}. 
Each pair was given a pair name by the researchers with which they used to login at a custom instance of \Scratch hosted at the University of Passau, which tracks the interactions with the code editor as well as the final program.

At the beginning of the course the students had to individually fill in a survey with their pair name, age, sex, previous knowledge in programming, if they know their pair partner, and if they think programming is cool. During the course, each student had to fill in the \textit{Fun Toolkit}~\cite{read2008} after each task~(\cref{tab:survey}), asking if they would do the task again (\textit{AG}), how they liked it (\textit{FU}), what their role was, and how they liked their role (\textit{LR}).
At the end of the course, the students were given stickers of different colors to mark the tasks they liked best and worst, and the ones they found easiest and most difficult. At the end, they were asked to answer the question of whether they think programming is cool again.

To ensure that the course ran smoothly, researchers filled different roles. One researcher was exclusively responsible for explaining the tasks and guiding through the course, while four to six supervisors observed the pairs and assisted with questions. One supervisor was usually responsible for two pairs.
During the course, the supervisors observed the behavior of each pair assigned to them on the evaluation sheet after each task. In total, they kept track of six different categories: compliance with roles, collaboration, communication, harmony, dispute and need for help (\cref{tab:survey}). In order to reduce gender biases in our setting, the role of the instructor and the supervisor were gender-balanced through all courses.

\subsection{Dataset}

A total of 139 children, 56 girls and 83 boys, from six different schools participated in the course. Of these, 32 students attended an elementary school and 107 a secondary school in the local area of Passau. The average age was 11.18 years (f: 11.16, m: 11.20) and most students (120) indicated that they knew their pair partner well (\textit{a little}: 12, \textit{not so well}: 7).  
More than half of the children stated that they had no prior programming experience at all (f: 35, m: 45), while 59 students had already programmed \textit{a little} (f: 14, m: 21) or even a little more (f: 7, m: 17).
In total there were 71 pairs, of which 66 were pairs of two (ff: 26, fm: 3, mm: 37). However, due to the absence of individual students in the classes, there were also groups of three and one male individual, which were excluded from the analysis. 

\subsection{Data Analysis}

To identify gender-dependent characteristics, our independent variable is the pair constellation: all-female, all-male and mixed pairs.
To determine the effects of this constellation on the dependent variables of attitude, behavior, and programming outcome, we consider the research questions as follows.

\subsubsection{\textbf{RQ1: Attitude}} \label{sec:methodRQ1}

To answer RQ1, we consider students' survey responses regarding their attitude towards programming and the course design, i.e., enjoyment and difficulty of the tasks and their assigned roles.  

\paragraph{Attitude Towards Programming}
To determine changes in attitude towards programming, we perform a pairwise comparison of whether students think programming is \textit{cool} before and after the course. 
We measure statistical differences using a Wilcoxon Rank Sum test with $\alpha\leq0.05$.

\paragraph{Attitude Towards Course Tasks}

To measure the overall hedonic quality of the tasks~\cite{read2008}, we sum the Likert values of the Again-Again-table \textit{AG} ([1,3]) and the Smileyometer \textit{FU} ([1,5]) for each student and task~(\cref{tab:survey}). The transformation of the Again-Again-table values corresponds to a value range of three gradations, where 1 is the lowest value (\textit{no}) and 3 is the highest value (\textit{yes}). The values of the Smileyometer correspond to a 5-point Likert scale, with the best value representing 5 and the lowest value representing 1.
For each student and task, we obtain a value in the interval [2,8], assuming they have completed all the tasks---otherwise, the interval is adapted to the completed tasks in order to be able to compare the different pair constellations. 
The sum of each pair is normalized to [0,1] with the following formula:
\begin{equation} \label{eq:1}
	 z_{i} = \frac{(x_{i}-min(x))}{(max(x)) - min(x))} 
\end{equation}
where $ x = (x_{i},...,x_{n})$ and $z_{i}$ is our $i^{th}$ normalized data. 
To compare the pair constellations and thus, determine gender-dependent effects, we sum the normalized values for each constellation and divide them by the number of students per constellation. Thus, we have a total value of retake and like per pair constellation across all tasks and per task. 

In addition, for each pair constellation, we rank the tasks that the students enjoyed the most and least based on their colored stickers. For this, we sum up the number of each sticker of a student for each task and pair constellation and provide the relative number per task. To identify which task the students considered easy or difficult, we use the same method for the stickers for easiest and most difficult task.

\paragraph{Attitude Towards PP Roles}
To determine whether the students preferred the role of driver or navigator, we consider the feedback from the Smileyometer \textit{LR}~(\cref{tab:survey}). Therefore, the sum of the values of each student at the Smileyometer are formed for both roles. 
Thus, two lists per student are created in the interval [3,15], although with potential deviations, since students may not adhere exactly to the compliance with roles and some tasks may not be processed due to technical problems. 
These interval values are again normalized to the range~[0,1] using \Cref{eq:1}.
After normalization, we compare the values within and between pair constellations for each task as well as for the respective role in total. We measure statistical differences using a Mann-Whitney-U Rank Sum test at a significance level of $\alpha\leq0.05$.

\subsubsection{\textbf{RQ2: Behavior}}
To answer RQ2, we consider the supervisors' survey responses from the survey regarding the six categories~(\cref{tab:survey}).
%
We sum the Likert values of each category ([1,5]) for each pair and task. For each pair and task, we obtain a value in the interval [6, 30] per category, assuming they have completed all the tasks---otherwise, the interval is adapted to the completed tasks. These interval values are normalized to the range [0,1] using \Cref{eq:1} in order to be able to compare the different pair constellations.
These normalized values are then aggregated for each pair constellation and divided by the number of pairs per constellation. Thus, we have six values for each pair constellation, one per category.
We again measure statistical differences between the pair constellations using a Mann-Whitney-U Rank Sum test at a significance level of $\alpha\leq0.05$.

\subsubsection{\textbf{RQ3: Code}}
To answer RQ3, we consider the actions performed during coding, and the resulting programs in terms of code metrics, quality, and creativity. 
To determine how students interact with the \Scratch environment while programming, we log all their interactions (\textit{events}) during the course.
To evaluate the code of the programs, we use the static analysis tool~\litterbox~\cite{fraser2021litterbox} to analyze the number and types of the block used, and complexity represented by an interprocedural version of \textit{cyclomatic complexity} (ICC) based on an interprocedural control-flow graph. 
To measure the quality of the programs, we determine poorly written code (\textit{code smells} or bugs)~\cite{fraser2021litterbox} and particularly well-written code (\textit{code perfumes})~\cite{obermuller2021}.  
To evaluate the creativity of the programs, we determine the sprites used and stages of the blocks that differ from our gender-neutral sample and which additional tasks they have chosen in both free tasks~(\cref{sec:task5}).

\subsection{Threats to Validity}
Since the course was embedded in an authentic school setting and not a free-time event, students may have behaved differently, although the teachers were not present for most of the course. In addition, gathering opinions from children within this age range is very challenging due to difficulties of the task and cognitive abilities~\cite{read2008,borgers2004}. We have chosen this age range to ensure a considerably number of participants since encouraging schools to participate is rather challenging in Germany, however, differences might occur due to these times of schooling.
Parents consented to use the data for research purposes, and no video or audio files of the students were recorded, except for approved images. The design of the study including the questionnaire is approved by a member of the ethics committee of the University of Passau. The course design is based on the established tutorial from the \Scratch website as well as several guidelines~\cite{williams2008,grover2015}, and independently evaluated by multiple didactic experts. Since the students were programming novices, the first four tasks were implemented with a high degree of guidance, although both free tasks were done independently with assistance only when needed. The observations of the supervisors are inherently subjective, however, all researchers have a background in computer science didactics and have already supervised numerous courses. To be able to verify all of our results and replicate the study, we provide all course materials and analyses online.

\section{Results}

As part of our experiment procedure, we explicitly refrained from
assigning the pairs in order to not bias the students and to provide a
realistic classroom setting. Since the resulting number of mixed pairs
is proportionally very low (n=3), it is unlikely to provide
representative and statistically significant results, and we therefore
focus on both same-sex pairs in the analysis, and only mention
specifics of the mixed pairs in the discussion where they provide
relevant insights. Clearly, one lesson learned in this context is
that, for better or for worse, students in this age group prefer to
work with the same sex.

\begin{table}
\centering
\caption{The normalized scores of how much students liked the task in the different pairs.}
\label{tab:funFactor}
\begin{tabular}{lrrrrrrr}
\toprule
 Pair & 1 & 2 & 3 & 4 & 5 & 6 & $\varnothing$\\
\midrule
ff & 0.86 & 0.79 & 0.83 & 0.76 & 0.80 & 0.87 & 0.82 \\
mm & 0.76 & 0.80 & 0.82 & 0.81 & 0.88 & 0.90 & 0.83 \\
\bottomrule
\end{tabular}
\end{table}
\begin{table}[t]
\centering
\caption{The normalized scores of how much students liked their role for each task in the different pairs.}
\label{tab:likeRole}
\begin{tabular}{llrrrrrrr}
\toprule
Role & Pair & 1 & 2 & 3 & 4 & 5 & 6 & $\varnothing$\\
\midrule
\multirow[c]{3}{*}{Driver} & ff & 0.77 & 0.77 & 0.84 & 0.73 & 0.79 & 0.88 & 0.80 \\
 & mm & 0.80 & 0.82 & 0.80 & 0.85 & 0.90 & 0.88 & 0.84 \\ \midrule
\multirow[c]{3}{*}{Navigator} & ff & 0.72 & 0.71 & 0.81 & 0.76 & 0.85 & 0.91 & 0.72  \\
 & mm& 0.72 & 0.78 & 0.79 & 0.86 & 0.86 & 0.88 & 0.75 \\
\bottomrule
\end{tabular}
\end{table}

\subsection{RQ1: Attitude}
\label{sec:rq1}

To determine the students' attitude towards programming and their roles, we consider the coolness factor, fun factor, rankings of difficulty and enjoyment, and evaluation of roles.


\subsubsection{Attitude Towards Programming}

\begin{figure*}[t]
\centering
{\includegraphics[width=0.75\textwidth]{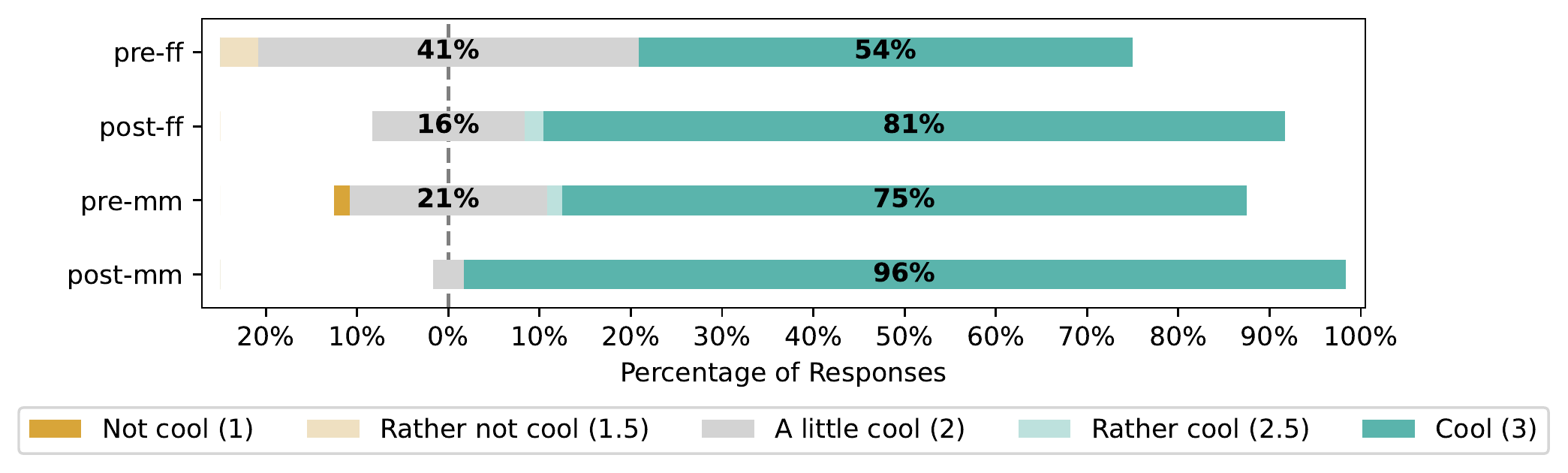}}
\caption{Attitude towards programming before/after the course.}
\label{fig:likertAttitude}
\end{figure*}

Since a positive attitude towards programming is essential and girls often have a negative one~\cite{zhan2015b, sun2022}, we determine the students' attitude towards programming from the start to the end of the course~(\Cref{fig:likertAttitude}). 
We observe a significant change for both same-sex pairs~(both $p < 0.001$).
Overall, an increase towards a positive attitude exists for both types of pairs, although the increase is largest for the all-female pairs~(\Cref{fig:likertAttitude}). Thus, it seems that our course design, in addition to the self-experience, contributes to avoid stereotypical preferences and thus, enables especially girls to change their mindset towards programming~\cite{zhan2015b}. 

\subsubsection{Attitude Towards Course Tasks}\label{sec:rankingAttitude}


Considering the results of the Fun Toolkit (\Cref{tab:funFactor}), all students liked the tasks and would do them again, and there are only minor differences between the all-female and all-male pairs. Thus, the design and implementation of the tasks seem well suited for this age group and universally applicable. We assume that potential differences arising in our analysis are thus more likely due to PP behavior rather than gender biases regarding the task.


\begin{figure}[t]
\centering
\subfloat[\label{fun} Ranking of fun.]{\includegraphics[width=\columnwidth]{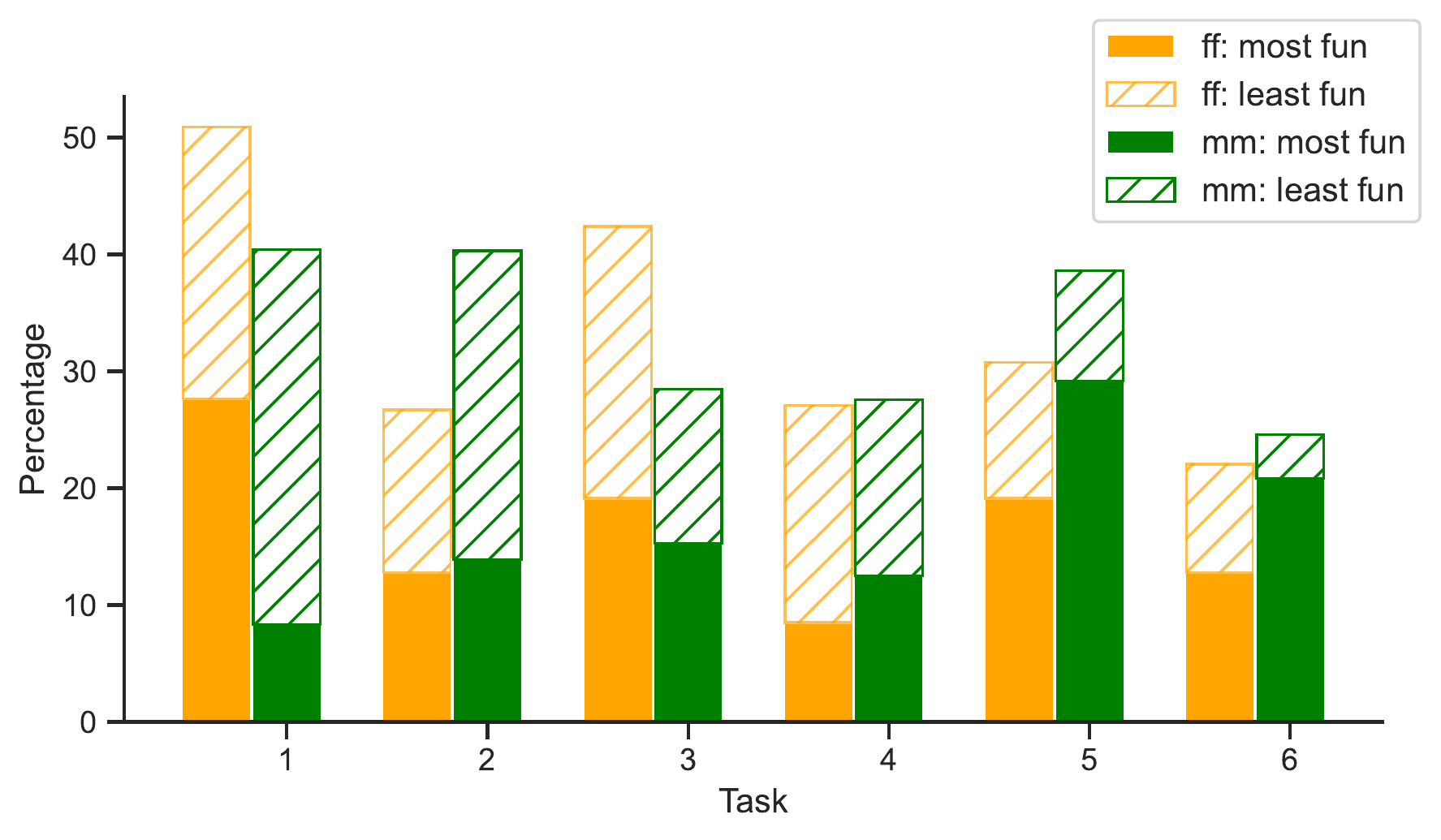}}
\quad
\subfloat[\label{difficulty} Ranking of difficulty.]{\includegraphics[width=\columnwidth]{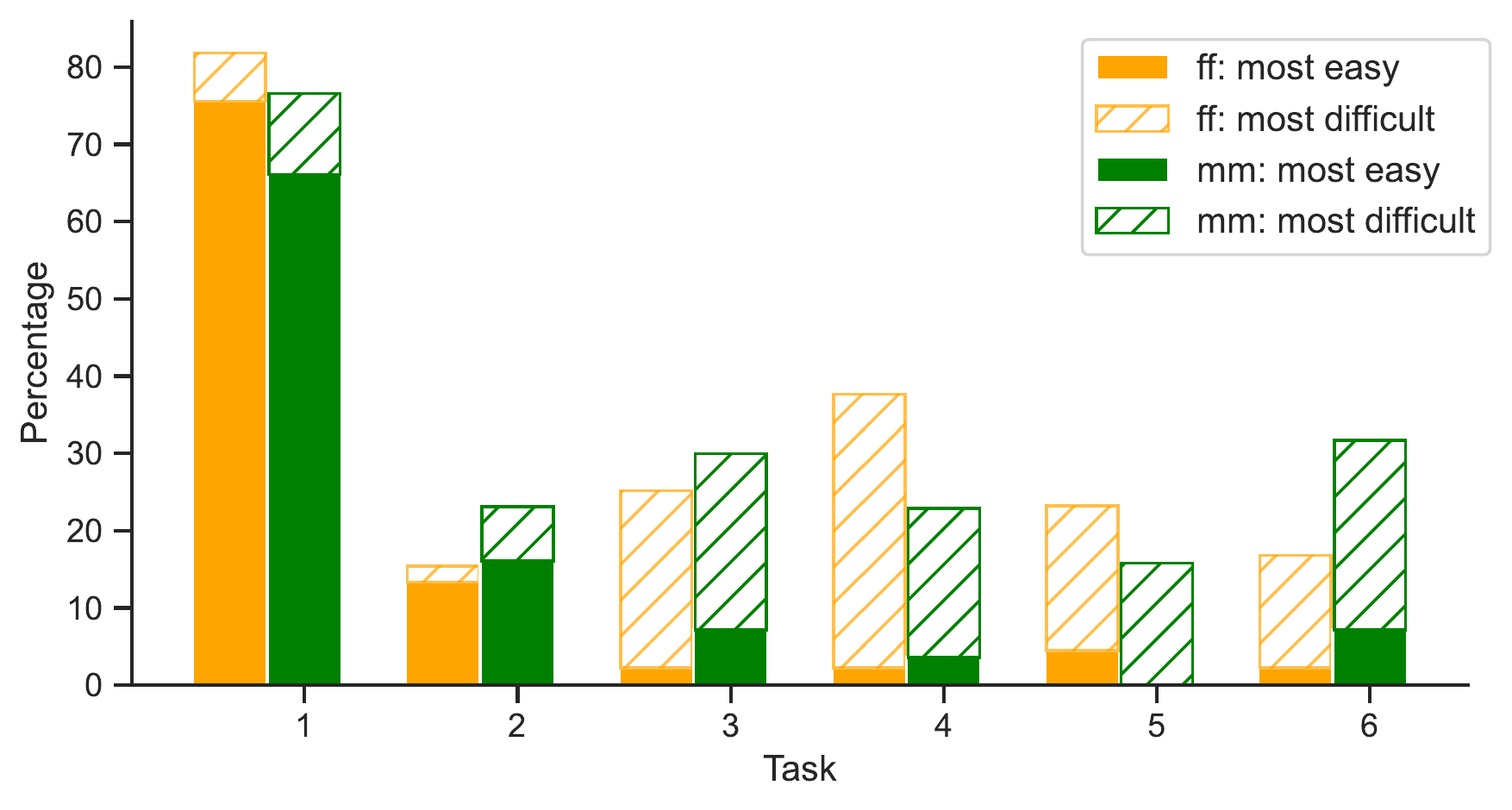}}
\caption{Ranking from students’ responses for each task regarding their perceived fun and difficulty in percentage.}  
\label{fig:ranking}
\end{figure}

\Cref{fig:ranking} shows the students' ranking of the tasks they liked best or least, and found easiest or most difficult (percentage of colored stickers, cf. \Cref{sec:methodRQ1}).
The rankings of fun and difficulty differ between the pairs, although not significantly.
While all-female pairs considered the first task to be the most enjoyable, for the all-males the fifth task was by far the best, and over one-third of the boys found the first task the least fun. 
This is intriguing, since both same-sex pairs agree that task one was the easiest by far. 
Thus, girls seem to like the simplicity or the playful and creative aspect of this task~\cite{grassl2021,grassl2022}, which makes it a suitable introductory task to lower the girls' initial hurdle, while boys seem to enjoy such kind of tasks less and may need an additional task, which results in a balancing act between having sufficient freedom for creativity versus solving a challenge~\cite{robertson2012}. 

While the all-female pairs experienced the fourth task as the most difficult by far, the all-male pairs found the last task the most challenging~(\Cref{fig:ranking}). 
The fourth task focused on the self-directed implementation of loops and conditionals, and the last task was the second free creative activity---so both same-sex pairs may have felt overwhelmed by these tasks due to their autonomy and demand for adaption~\cite{benade2017}.
In particular, since the fifth task was also a free task, which is also the most popular among the young boys~(\Cref{fig:ranking}), the question emerges as to why task 6 is perceived as more difficult by the boys. We conjecture that the boys either expanded their projects creatively during the fifth task to such an extent that (1) they could not think of anything else to do and therefore found it tough or ``boring'', or (2) ``destroyed'' the project more during task 5, thus inhibiting task 6.
However, both same-sex pairs considered the third task to be the second most difficult, with about 23 \% each~(\Cref{fig:ranking}). Thus, the introduction of conditionals seems to be universally challenging~\cite{grover2015, adams2012}. 


\subsubsection{Attitude Towards PP Roles}


To determine which PP roles the students enjoyed the most, we consider how much the pairs liked their assigned roles (\Cref{tab:likeRole}). 
Overall, students in all pairs found the role of the driver to be significantly better than that of the navigator (ff: $p = 0.011$, fm: $p = 0.015$, mm: $p = 0.028$) (\Cref{tab:likeRole}). The boys disliked the navigator role the most in the first task, suggesting they probably also wanted to take the lead here. 
However, the all-female pairs preferred the navigator slightly from the fourth task to the last. This suggests that the girls prefer not to be in charge of controlling the computer the more self-reliant the task becomes. 
%
Compared across tasks, all-female pairs were most comfortable with either role in the last task, and the boys similarly stated to like both roles the most in the last two tasks, which is likely because they had become familiar to the roles~\cite{williams2002a}.


Comparing the pairs shows that mixed pairs enjoy the role of the navigator significantly less across all tasks than both same-sex pairs ($\varnothing\;0.47$, both $p < 0.001$). 
We observe one single statistically significant difference for the driver between the two same-sex pairs in task five, where the all-male pairs enjoyed the task more than the all-female pairs ($p = 0.037$).
%
%
%


\rqsummary{RQ1}{All students had a highly positive attitude towards the course design and the roles of PP, but preferred being the driver. All students considered programming after the course to be cooler. The perceived fun and difficulty per task differs between all-female and all-male pairs.}

\subsection{RQ2: Behavior}
\label{sec:rq2}

\begin{table}
\centering
\caption{Pair behavior from the supervisors' observations.}
\label{tab:supervisors}
\resizebox{\columnwidth}{!}{%
\begin{tabular}{llrrrrrrr}
\toprule
Category & Pair & 1 & 2 & 3 & 4 & 5 & 6 & $\varnothing$ \\
\midrule
\multirow[c]{3}{*}{Compliance} & ff & 0.91 & 0.78 & 0.85 & 0.73 & 0.93 & 0.84 & 0.84 \\
 & mm & 0.84 & 0.76 & 0.76 & 0.79 & 0.63 & 0.68 & 0.76 \\  \midrule
\multirow[c]{3}{*}{Collab.} & ff & 0.79 & 0.85 & 0.90 & 0.83 & 0.86 & 0.91 & 0.84 \\
 & mm & 0.85 & 0.83 & 0.80 & 0.82 & 0.85 & 0.81 & 0.83 \\ \midrule
 \multirow[c]{3}{*}{Commun.} & ff & 0.68 & 0.76 & 0.85 & 0.79 & 0.82 & 0.93 & 0.78 \\
 & mm & 0.83 & 0.84 & 0.85 & 0.88 & 0.88 & 0.86 & 0.85 \\ \midrule
 \multirow[c]{3}{*}{Dispute} & ff & 0.06 & 0.04 & 0.00 & 0.04 & 0.00 & 0.00 & 0.03 \\
 & mm & 0.06 & 0.09 & 0.12 & 0.02 & 0.18 & 0.12 & 0.10 \\ \midrule
\multirow[c]{3}{*}{Harmony} & ff & 0.82 & 0.87 & 0.93 & 0.90 & 0.95 & 0.98 & 0.89 \\
 & mm & 0.92 & 0.88 & 0.87 & 0.87 & 0.88 & 0.87 & 0.88 \\ \midrule
\multirow[c]{3}{*}{Help} & ff & 0.29 & 0.26 & 0.25 & 0.60 & 0.70 & 0.52 & 0.41 \\
 & mm & 0.11 & 0.21 & 0.36 & 0.42 & 0.38 & 0.50 & 0.32 \\
\bottomrule
\end{tabular}%
}
\end{table}


To determine the behavior of students in their pairs during the course, we analyze the observations of the supervisors within six categories.
\Cref{tab:supervisors} shows the normalized values of these six observed categories for each pair per task.


\subsubsection{Compliance}

\begin{figure}[tb]
\centering
{\includegraphics[width=\columnwidth]{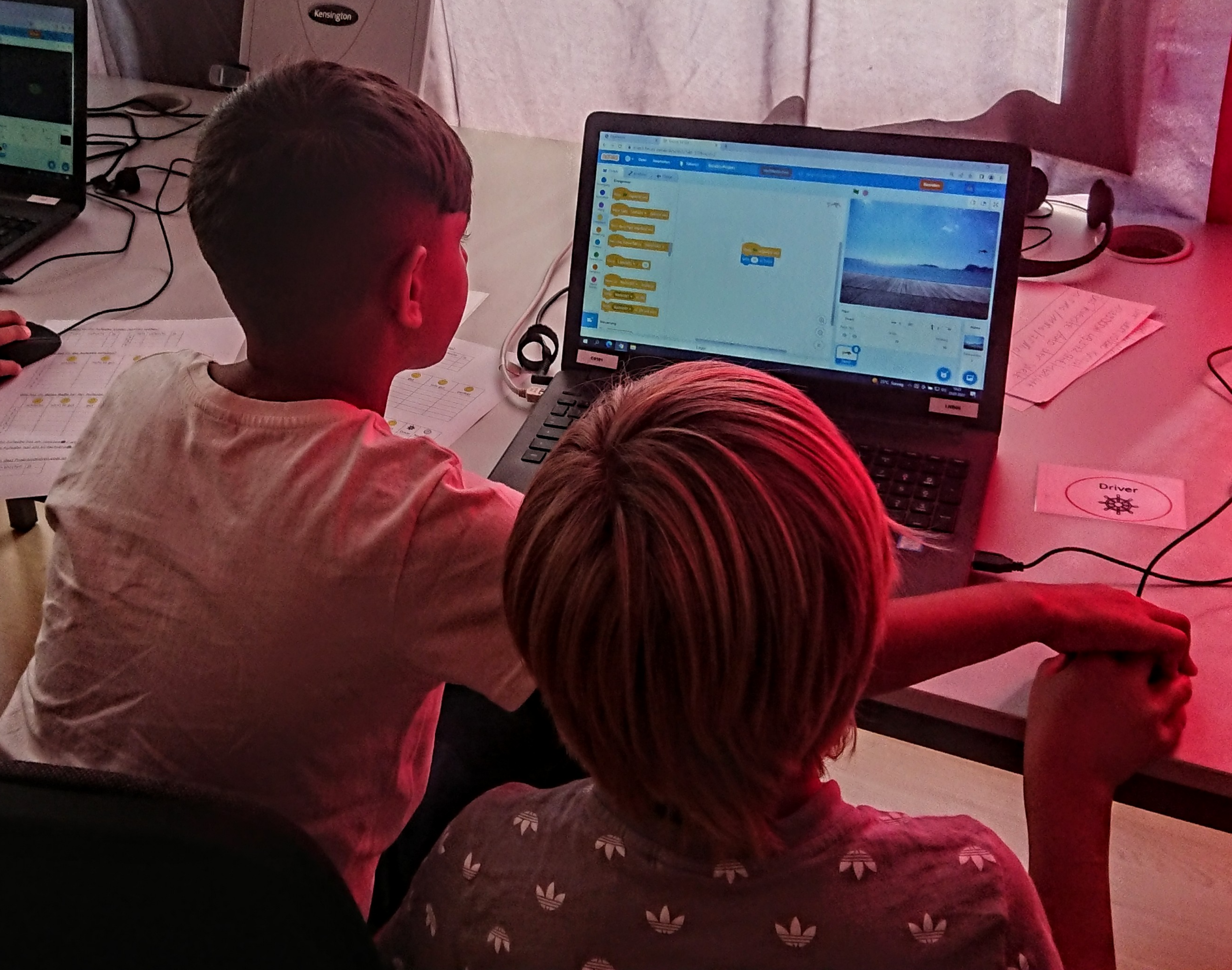}}
\caption{One boy interferes with the role distribution by reaching for the mouse although he is navigator.}
\label{fig:boys}
\end{figure}

In terms of how well the pairs followed their assigned roles (\textit{CP}), we observe marginal differences between both same-sex pairs ($p=0.332$), with the all-female pair following them slightly more in each task~(\Cref{tab:supervisors}). However, we notice a significant difference in the fifth task ($p=0.031$), where the all-male pairs followed their roles considerably less than the all-female pairs~(\Cref{tab:supervisors}). This may be due to the subject of the fifth task, where the students were given creative freedom to implement their own ideas, and where both boys of a team seemed to want to assert themselves~(\Cref{fig:boys})~\cite{wieselmann2020a,campe2020}. Although not significant, a similar behavior can also be observed in the last task, 
which is the second offhanded creative task, and thus supports the theory of the assertion within all-male pairs.
%
%
%
Overall, these results confirm prior observations that girls stick to instructions more often than boys due to socially expected behavior~\cite{chaplin2013,harskamp2008,pomerantz2001,sullivan2019}, and this generalizes to PP.

\subsubsection{Collaboration}\label{sec:collab}
Collaboration is essential for successful teamwork~\cite{denner2021,campe2020}. Both same-sex pairs consistently collaborated very well ($p=0.608$)~(\Cref{tab:supervisors}). 
While in the all-female pairs the first task is salient, possibly due to differences of opinion in the choice of sprites, in the all-male pairs almost all tasks are affected by outliers, especially for the last three tasks, where they were given greater freedom regarding the implementation of the tasks~(cf. \Cref{sec:dispute}). 
However, the mixed pairs show a comparatively low level of collaboration ($\varnothing\;0.52$), which may indicate conflicts between the sexes~\cite{wieselmann2020a}.

\subsubsection{Communication}
Communication is one of the key factors for successful teamwork. Analogously to the collaboration, we see little difference between the same-sex pairs ($p= 0.379$)~(\Cref{tab:supervisors}), but with a large difference to the mixed pairs ($\varnothing\;0.42$). However, the all-female pairs communicate somewhat worse than the all-male pairs
in the first and last tasks. 
In particular, in the first task, the distribution of both pairs is different, which has already been observed in the collaboration~(cf. \Cref{sec:collab}). Noticeably, the behavior seems to almost reverse in the last two tasks, where first the all-female pairs show a higher variance in communication than the all-male pairs, which might be a result of the boys finding the task difficult~(cf. \Cref{sec:rankingAttitude}).

\subsubsection{Dispute}\label{sec:dispute}

When it comes to disputes between the pair members, we observe several differences: all-female pairs show the lowest conflict rates in total, while all-male pairs show a high potential of conflict~(\Cref{tab:supervisors}); the mixed pairs show the highest number of disputes ($\varnothing\;0.20$). 
In particular, the tasks three, five and six led to most conflicts among the all-male pairs, 
with significant differences between the all-female and all-male pairs for the third ($p=0.025$) and fifth task ($p=0.022$). This high conflict rate  might be due to the demand for leadership, even at this young age~\cite{harskamp2008, campe2020, denner2021, wieselmann2020a}. This is especially reflected in the fifth task, which has the lowest level of compliance, yet, the highest conflict rate among the all-male pairs, suggesting that the interference of assigned roles leads to a power struggle~(\Cref{fig:boys}).
Regarding the all-female pairs, the first task led mainly to conflict~(\Cref{tab:supervisors}). Based on our observations, this might be due to the decision making when picking a sprite and background.

\subsubsection{Harmony}

Similar to the conflict rate, we see comparable values in the category to what extent the pairs dealt with each other harmoniously: likewise, the mixed pairs ($\varnothing\;0.57$) are far below the same-sex pairs, which were comparably harmonious across all tasks~(\Cref{tab:supervisors}). In-line with previous findings on the level of conflict, we observe a decrease in harmony in the first task in the all-female pairs, whereas in the remaining tasks the boys are noticeably more inharmonious.

\subsubsection{Help}\label{sec:help}
Noticeably, we identify a wide range of help sought by both same-sex pairs ($p=0.105$).
Overall, the all-female pairs sought more help than the all-male pairs~(\Cref{tab:supervisors}), which in particular is significant for task five ($p=0.023$). Since this is the first task in which they implement own ideas, the all-female pairs might have needed guidance regarding the implementation~\cite{benade2017}, whereas the all-male pairs seemed to prefer to try it out for themselves.

\rqsummary{RQ2}{Overall, the behavior of the young students shows that PP works very well with its roles, and both same-sex pairs cooperated and communicated very well. The all-male pairs showed greater potential of interfering roles and dispute, while all-female pairs sought slightly more help.}
 \subsection{RQ3: Code}
\label{sec:rq3}

To evaluate the impact of the different pairs on the students' programs, we consider the interface events, code metrics, quality, and creativity of the programs. 



\begin{figure}[tb]
\centering
\includegraphics[width=\columnwidth]{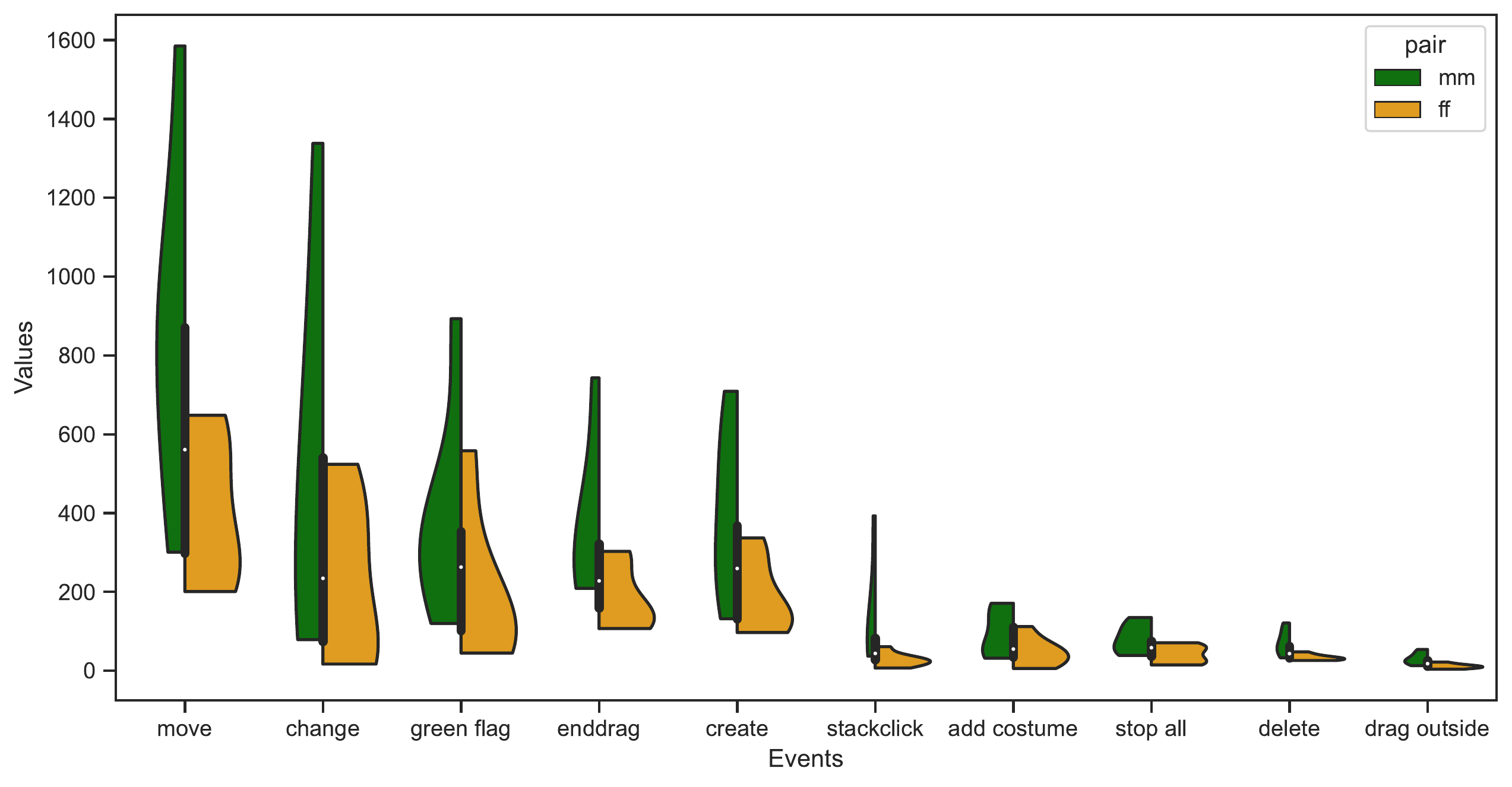}
\caption{Distribution of events over all tasks.}  
\label{fig:events}
\end{figure}

\subsubsection{Interaction and Events}
While creating code, the all-male pairs interacted more with the \Scratch interface than the all-female pairs (\Cref{fig:events}).
%
In particular, the all-male pairs moved blocks around significantly more often than all-female pairs ($p=0.041$). 
Similarly, boys changed the parameters of an existing block more than the girls ($p=0.060$), for instance to change the number of steps of a sprite (\Cref{fig:events}). Dragging a new block out of the toolbox is also done more often by the all-male pairs, but not with such a distinct difference ($p=0.093$). Comparatively few blocks were deleted or pulled out of range, yet, the boys removed blocks significantly more frequently ($p=0.019, p=0.015$) (\Cref{fig:events}). Overall, boys seem to like to modify their programs. However, this difference is only noticeable for code; both same-sex pairs similarly added new costumes to their sprites ($p=0.309$).

Besides the modifications, we also note that all-male pairs execute stack clicks (double-clicking blocks or scripts leading to their execution) significantly more often than the all-female pairs ($p=0.008$).
Clicking the green flag to start the entire program ($p=0.092$) is also more often done by the all-male pairs, suggesting that they run their programs more frequently.

The high number of stack clicks and changes suggests that the all-male pairs seem to be more adventurous, as one probably clicks on scripts to try out things (``what happens if we do this'') or fixing their programs as they might be more prone to introduce code smells~(\Cref{tab:smells}) due to their experimentations. In a sense \Scratch is built to foster this~\cite{resnick2009a,roque2016a}, however, in our PP setting this seems to be less appealing to girls, which might be either due to their preference to stick to instructions, or an aversion to experimenting. We also saw this kind of gender-dependent behavior in RQ2~(\Cref{sec:help}), where we observed that boys needed the least and girls the most help in the first creative task (task 5).


\begin{figure}[tb]
\centering
\includegraphics[width=\columnwidth]{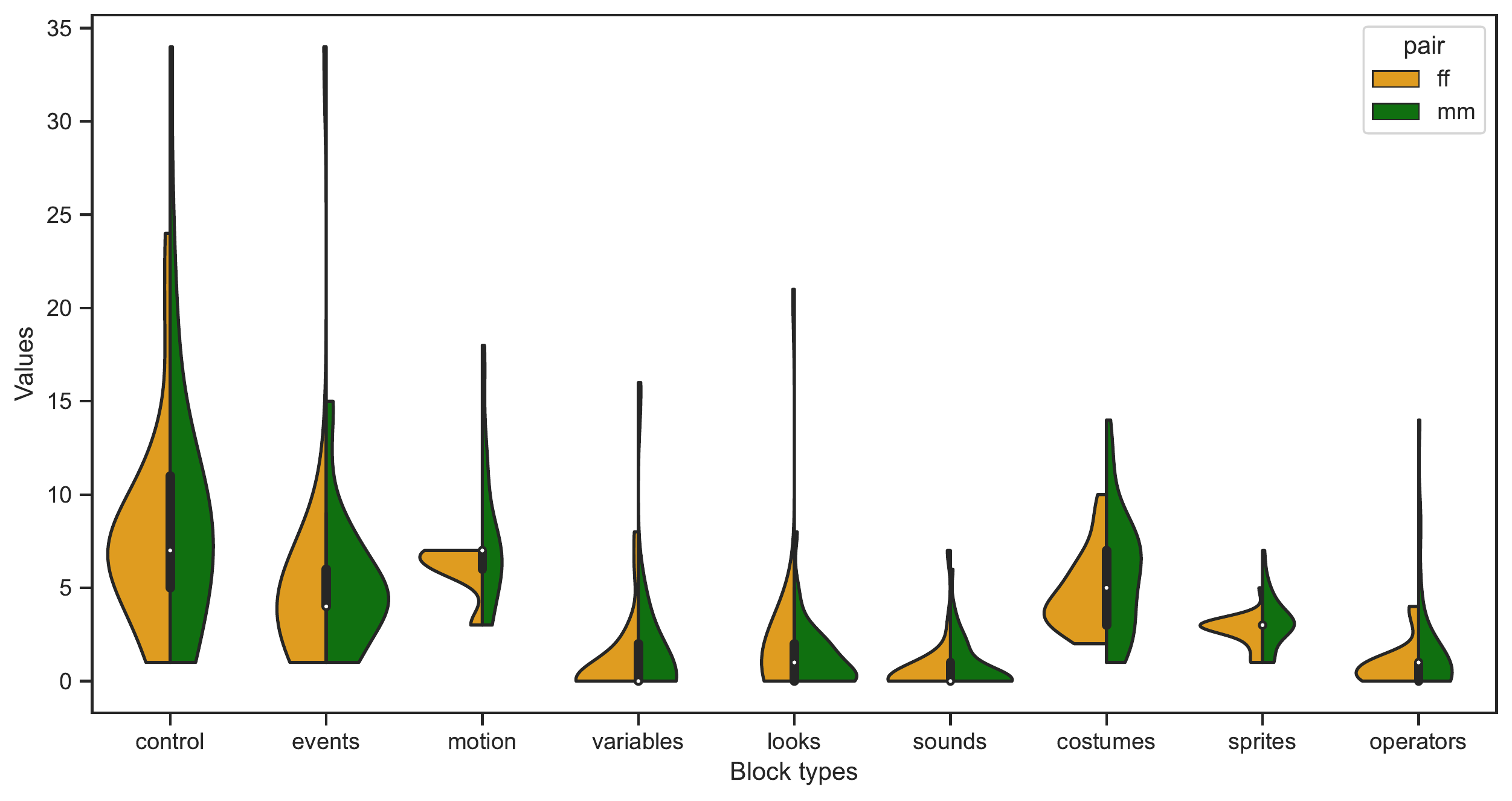}
\caption{Distribution of block types of the final projects.}  
\label{fig:blocktypes}
\end{figure}

\subsubsection{Use of Block Types}
The pairs used a similar number of blocks in total ($p=0.505$), although the distribution among the different block types differs~(\Cref{fig:blocktypes}).
Due to the mechanics of the game, \textit{control} and \textit{event} blocks are the most common, as they can be used to implement game-typical interactive elements. This design was explicitly chosen because prior work showed that girls, when given the freedom to do as they like, tend to implement stories that mostly use only sequences, and thus apply less programming knowledge~\cite{adams2012, grassl2021}.
Regarding the use of \textit{control} and \textit{event} blocks, there are few differences between both same-sex pairs in the final project ($p=0.526, p=0.310$), with slightly more \textit{control} blocks in the male ones.
\textit{Sound} blocks were similarly rarely used by all pairs.

However, we observe slight differences in the remaining categories:
One notable difference is the upper limit of \textit{motion} blocks in the all-female pairs, whereas the variance is higher in the all-male pairs. This is in-line with previous research reporting that girls use fewer \textit{motion} blocks than boys, and prefer \textit{looks} blocks instead~\cite{adams2012, grassl2021}. 
Although the project category of game~\cite{issaee2021} reduces this effect, it is nevertheless evident in the analyzed projects.


\subsubsection{Code Quality}

\begin{table}[t]
\centering
\caption{Mean code smells and perfumes per final project.}
\label{tab:smells}
\begin{tabular}{llrrr}
\toprule
Category & & ff & mm & $p$-value\\
\midrule
Code perfumes&conditional inside loop & 2.25  & 3.03 & 0.715 \\
&loop sensing & 1.96  & 2.57 & 0.463\\
&collision & 1.00  & 1.22 & 0.765\\
&initialisation of position & 0.50  & 0.68& 0.698\\
&boolean expression & 0.21  & 0.46 & 0.384\\
&backdrop switch & 0.67  & 0.00& 0.227\\
&initialisation of looks & 0.38  & 0.14 & 0.077\\
&custom block usage & 0.00 & 0.32 & 0.440\\ \midrule

Code smells &missing initialization & 0.92 & 1.86 & \bf{0.007}\\
&busy waiting & 0.62  & 0.59& 0.753 \\
&dead code & 0.29  & 0.54& 0.315\\ 
&empty sprite & 0.25 & 0.41& 0.355\\
&empty control body & 0.17  & 0.32& 0.383\\
&sprite naming & 0.04  & 0.32 & 0.093\\
&unnecessary if & 0.00  & 0.27& 0.161\\
&unused variables & 0.04  & 0.16& 0.235\\
\bottomrule
\end{tabular}
\end{table}

\Cref{tab:smells} shows the most prominent code smells and perfumes of the final projects.
%
%
Code perfumes~\cite{obermuller2021} describe recurring code patterns demonstrating the correct application of programming concepts.
The code perfume \textit{conditional inside loop} (i.e., a loop construct containing an if-condition) is the most prominent of all pairs per project. This advanced code structure is introduced in the third task (\Cref{sec:task3}), and although this is not reflected in the code perfumes, we observed that during their own application in task four (\Cref{sec:task4}) especially the girls needed a lot of help to implement exactly this construct. 
%
Noteworthy, the girls coded backdrop changes by implementing \textit{switch backdrop to} with event handler \textit{when backdrop switches to}, which the boys did not implement at all ($p=0.227$).
All-female pairs also implemented \textit{initialisation of looks} more frequently ($p=0.077$), i.e., \textit{look} blocks being present in \textit{when green flag clicked}, which suggests that they were trying to solve a subtask of the \textit{defining a start state problem}~\cite{obermuller2021}.
This solution strategy and the pattern \textit{initialization of postions} that occurs in both same-sex pairs ($p=0.698$) are particularly relevant for computational thinking skills~\cite{seiter2013}. 
Interestingly, all-male pairs implemented \textit{boolean expressions} twice as often as all-female pairs, although the difference is not statistically significant ($p=0.077$, \Cref{tab:smells}). Since all students had to use operators at least once in task four (\Cref{sec:task4}), this suggests that the boys used more operators in the free tasks.
%

Regarding code smells, \textit{missing initialization} is the most common bad code snippet in both pair constellations \Cref{tab:smells}, but it appears significantly more often in the all-male pairs than in the all-female pairs ($p=0.007$), which is in-line with prior research on gender-differences in programs created by individuals~\cite{grassl2021}. In-line with our observation on boys being more adventurous, this smell occurs when code is created that modifies different properties of sprites without properly initializing them.
%
We also note that boys tend to not name their sprites and just retain their default names (e.g., Sprite1) (\textit{sprite naming}, $p=0.093$), which is also known by prior research on individuals' programs~\cite{grassl2021,grassl2022}.


\begin{figure}[tb]
\centering
\includegraphics[width=\columnwidth]{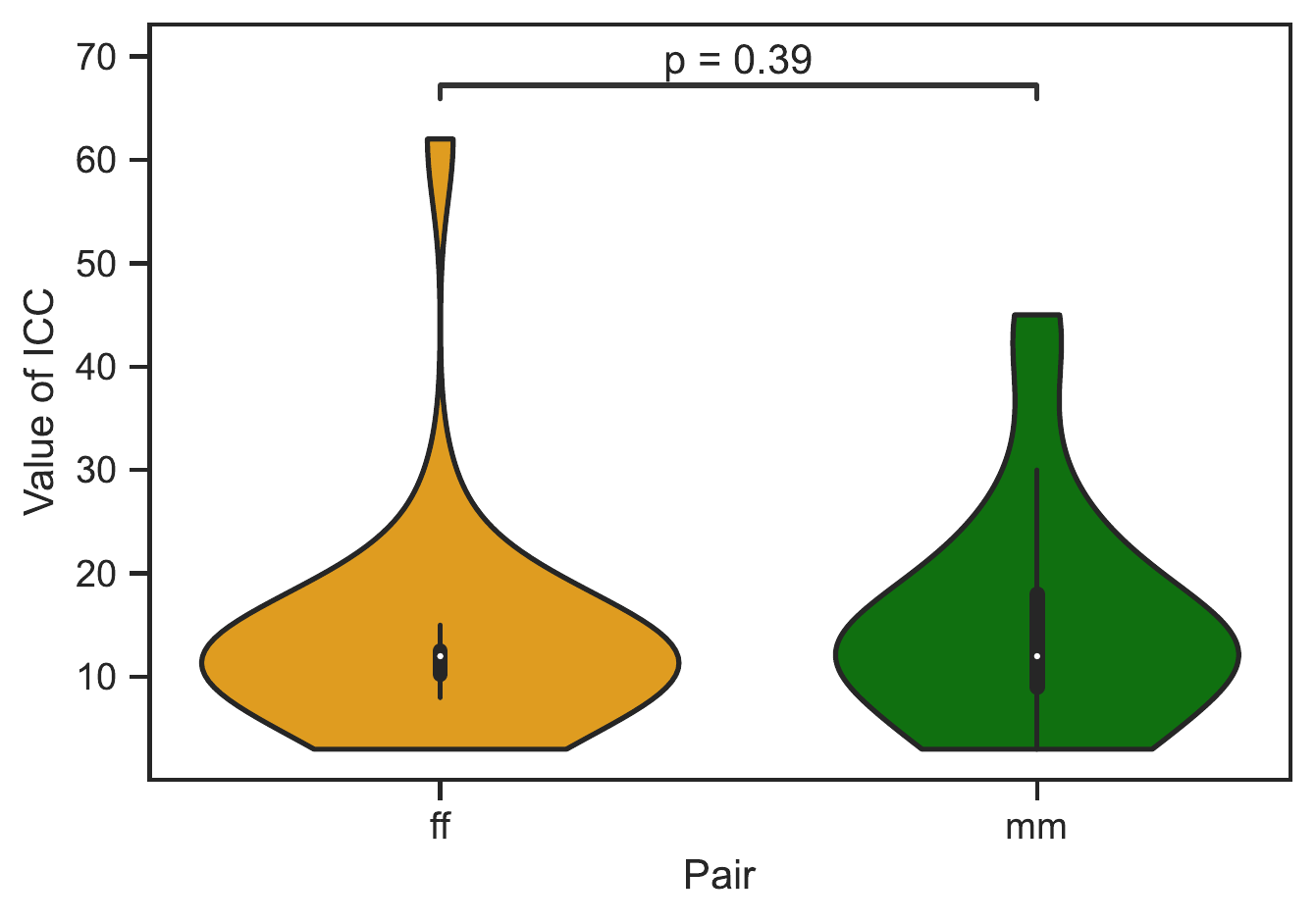}
\caption{Distribution of ICC of the final projects.}  
\label{fig:icc}
\end{figure}

\subsubsection{Code Complexity}
Regarding the code complexity of the final programs, we determine a rather similarly distributed complexity among the all-female and all-male pairs ($p = 0.390$), with the boys having slightly more programs with stronger complexity~(\Cref{fig:icc}). This is likely due to the increased number of interactions they implemented in the game, which can also be observed in the control blocks~(\Cref{fig:blocktypes}).
However, it is notable that the most complex project was implemented by a female pair ($ICC = 62$,~\Cref{fig:icc}), in which not only scores are awarded, but which also introduces different levels, represented by several variables and stage changes. However, even when excluding this outlier there are no significant differences ($p=0.240$).
Overall, the projects are comparatively complex for an introductory programming project, which we explicitly encouraged through the course design (e.g., game rather than story) in order to improve the girls' adaptation to programming concepts~\cite{grassl2021,robertson2012,aivaloglou2017}. 


\begin{figure}[tb]
\centering
\subfloat[\label{sprites-ff} Sprite names of all-female pairs.]{\includegraphics[width=\columnwidth]{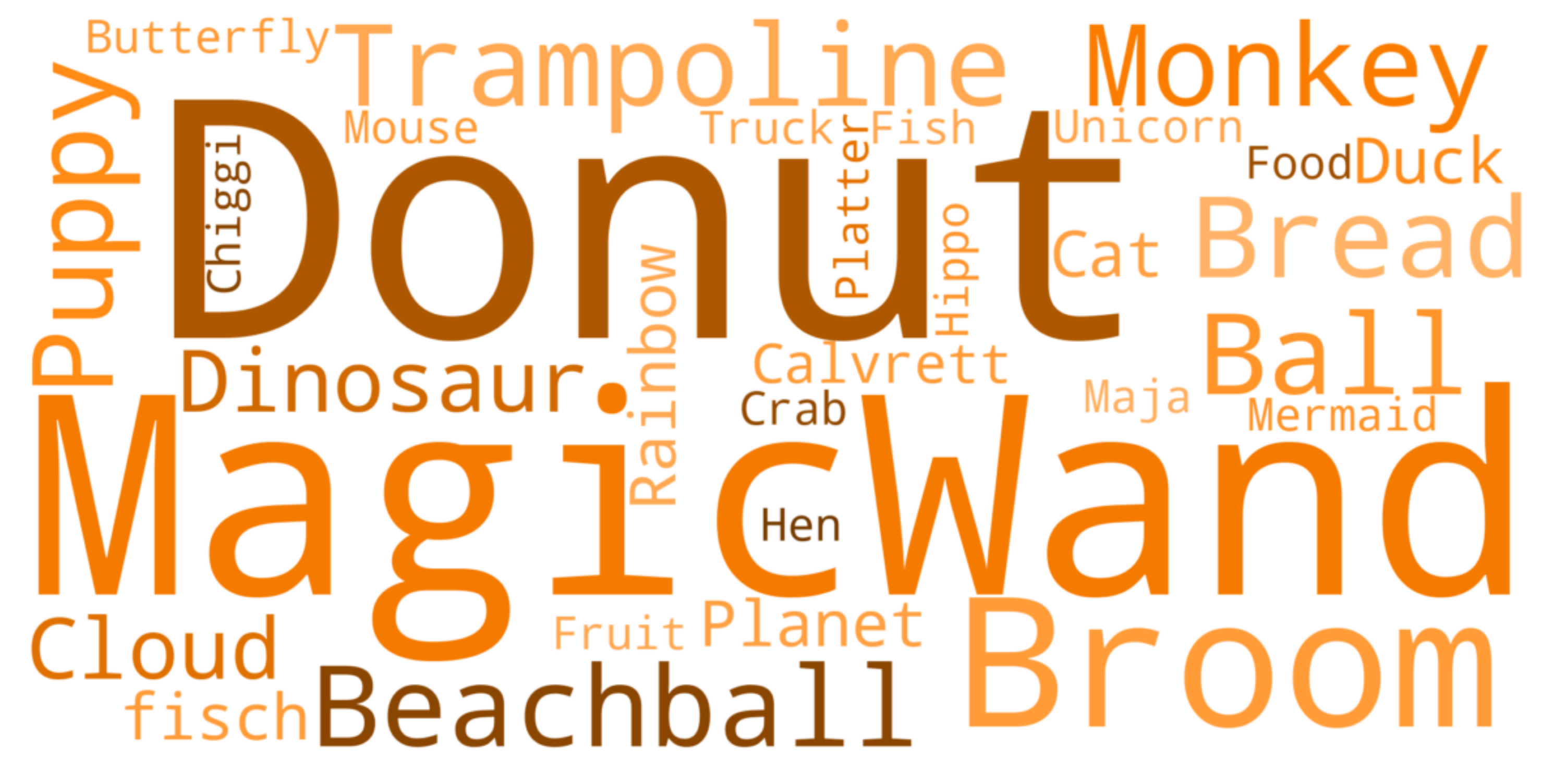}}
\quad
\subfloat[\label{sprites-mm} Sprite names of all-male pairs.]{\includegraphics[width=\columnwidth]{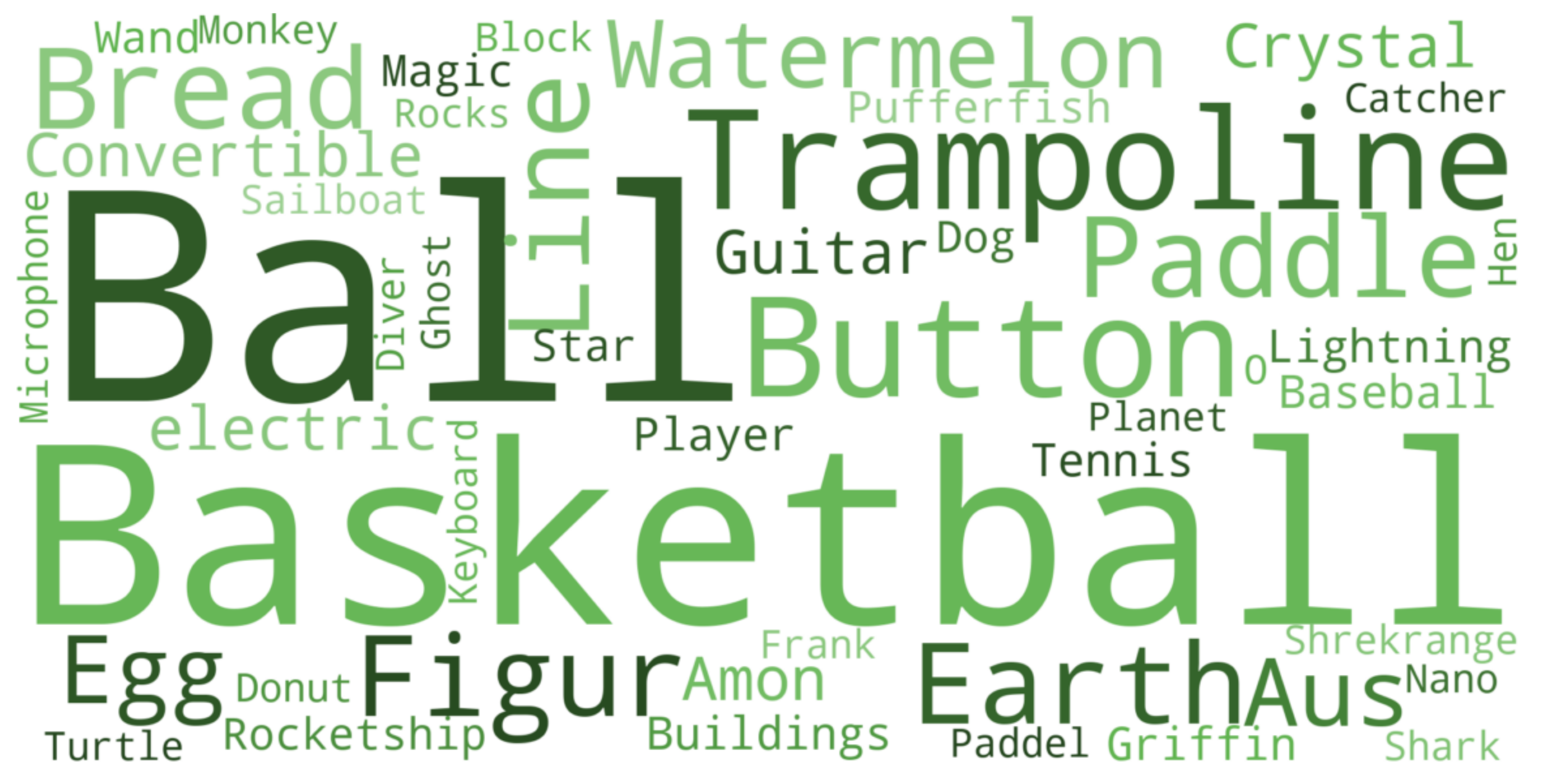}}
\caption{Distribution of the names of the self-chosen sprites, where the size of the name indicates its frequency.}  
\label{fig:wordclouds}
\end{figure}

\subsubsection{Creativity}

\Scratch enables young students to express their creativity while programming, which supports learning and motivational success~\cite{roque2016a}. We consider the choice of sprites as an indicator of creativity, as this enables students to adapt their programming to their own ideas~\cite{robertson2012}.
\Cref{fig:wordclouds} summarizes the sprite preferences for the implementation of the pong game as word clouds.
In particular, the sprite \textit{ball} (ff: 8.33~\%, mm: 21.62~\%) followed by the \textit{paddle} (ff: 25.0~\%, mm: 29.72~\%) were implemented less by the all-female pairs than the all-male pairs, whereas the girls used the \textit{line} more often (ff: 62.5~\%, mm: 56.75~\%).

\Cref{sprites-mm} shows the boys' preference for the sprite \textit{ball} as well as other ball sports such as \textit{basketball} and \textit{baseball}, which have gender-stereotypical male associations. In contrast, the \textit{donut} and \textit{magic wand} are most popular among females~(\Cref{sprites-ff}). In addition, \textit{broom} seems to be a popular character, which in combination with \textit{magic wand} suggests magical worlds. However, besides gender stereotypical sprites like \textit{rainbow} or \textit{unicorn}, we also encounter neutral sprites like \textit{puppy}, \textit{monkey} or \textit{beachball}.
Furthermore, \textit{trampoline} and \textit{bread} are common to both pair constellations--these figures are well suited as an alternative for the \textit{paddle} due to their shape~(\Cref{fig:wordclouds}). 
Thus, neither our neutral course design nor the PP protocol inhibited either type of same-sex pairs from incorporating their own creative preferences. Those partly universally, but mostly socially learned gender-stereotypical preferences are in line with prior research~\cite{robertson2012,grassl2021,grassl2022}.

\rqsummary{RQ3}{All-male pairs interact considerably more with the \Scratch interface, and the use of block types differ between both pairs. However, the code quality and complexity is comparable. Regardless of the PP protocol, both same-sex pairs implement their own, partly gender-stereotypical, creative preferences.}
\section{Conclusions and Future Work}
\label{conclusions}

Pair programming is not only useful in practice but also in
programming education. One of the concerns in programming education is
how to better engage young female learners. It is therefore important
to understand whether gender-specific differences can be observed with
young learners performing pair programming. In order to investigate
this question, we designed an in-class introductory programming course
for young programming learners based on \Scratch, in which pair
programming is integrated, and studied it on 139 students aged between
8 and 14 years.

While the attitude towards programming and the course design is
overall positive, the perception of fun and difficulty of programming
tasks differ between all-female and all-male pairs. Both 
constellations prefer the role of the driver, although all-male pairs
do not adhere to their roles as well as girls. In addition, we
observed that all-male pairs are more adventurous in exploring the
possibilities of \Scratch and experimenting, while pairs of
girls stick to the instructions. This suggests socially learned
behavior which is transmitted to programming within same-sex
pairs. However, unlike prior studies on gender differences for
individual young programmers, we observe only minor differences in
code quality and complexity of the final programs among the
pairs, indicating that PP narrows this gap. When given creative freedom to the pairs, however, even pairs
fall back to gender stereotypical preferences.

An individual study like ours can only provide initial insights, but
raises several avenues for future research:
\begin{itemize}
\item We deliberately used a gender-neutral design, which raises the question how results would differ when using common stereotypical designs.
\item Our study consists of a relatively short programming session, which raises the question how results would differ if PP were applied throughout a longer period.
\item While our course design provides an easily reproducible PP scenario, deploying PP in the classroom will require research on dedicated teacher training.
\item Since it may be the case that gender-dependent differences become more pronounced with age, it would be interesting to replicate our study with higher grades.
\item Since online teaching has taken a more prominent role in education since Covid, it would similarly be interesting to investigate gender-specific effects in an online setting.
\end{itemize}
To support this future research we provide all course materials and
evaluations for replication online.

%

\section*{Acknowledgements}\label{sec:acknowledgements}
This work is supported by the Federal Ministry of Education and Research
through project ``primary::programming'' (01JA2021) as
part of the ``Qualitätsoffensive Lehrerbildung'', a joint initiative of the
Federal Government and the Länder. The authors are responsible for the content
of this publication.

\balance

\bibliographystyle{IEEEtran}
\bibliography{references}
\end{document}